\documentclass[superscriptaddress,floatfix,twocolumn,showpacs,preprintnumbers,amsmath,amssymb,prl]{revtex4-1}

\usepackage{graphicx}
\usepackage{dcolumn}
\usepackage{bm}
\usepackage{color}

\begin{document}


\title{Low-Temperature Low-Field Phases of the Pyrochlore Quantum Magnet Tb$_2$Ti$_2$O$_7$}

\author{L. Yin}
\email{yin@phys.ufl.edu}
\author{J. S. Xia}
\author{Y. Takano}
\author{N. S. Sullivan}
\affiliation{National High Magnetic Field Laboratory and Department of Physics, University of Florida, Gainesville, FL 32611, USA}
\author{Q. J. Li}
\author{X. F. Sun}
\email{xfsun@ustc.edu.cn}
\affiliation{Hefei National Laboratory for Physical Sciences at Microscale, University of Science and Technology of China, Hefei, Anhui 230026, People's Republic of China}%

\date{\today}

\begin{abstract}
By means of ac magnetic-susceptibility measurements, we find evidence for a new magnetic phase of Tb$_2$Ti$_2$O$_7$ below about 140\,mK in zero magnetic field. In magnetic fields parallel to [111], this phase is characterized by frequency- and amplitude-dependent susceptibility and extremely slow spin dynamics. In the zero-temperature limit, it extends to about 67\,mT (the internal field $H_{int}$\,$\simeq$\,52\,mT), at which it makes transition to another phase. The field dependence of the susceptibility of this second phase, which extends to about 0.60\,T ($H_{int}$\,$\simeq$\,0.54\,T) in the zero-temperature limit, indicates the presence of a weak magnetization plateau below about 50\,mK, as has been predicted by a single-tetrahedron four-spin model, suggesting that the second phase is a quantum kagome ice.

\end{abstract}


\maketitle


In rare-earth-titanate pyrochlores, R$_2$Ti$_2$O$_7$, trivalent rare-earth ions R$^{3+}$ with eightfold oxygen coordination form a three-dimensional lattice of corner-sharing tetrahedra. Alternatively, these magnetic oxides can be viewed as kagome layers of rare-earth spins coupled via interspacing triangular lattices of rare-earth spins, stacked along the [111] direction. In either view, the geometry leads to frustrated nearest-neighbor exchange interactions, whose interplay with an anisotropy dictated by a local $<$111$>$ direction, dipole-dipole interaction, and quantum fluctuations in some cases, result in various exotic magnetic ground states \cite{gingras2}. Dy$_2$Ti$_2$O$_7$ and Ho$_2$Ti$_2$O$_7$ are spin ices of classical Ising spins \cite{ram,morris,bram,harris}. By contrast, Gd$_2$Ti$_2$O$_7$ and Er$_2$Ti$_2$O$_7$ undergo magnetic transitions at temperatures of the order of 1\,K, but their ground states are unconventional, with fluctuations persisting down to 20\,mK \cite{yao05,lago05}. The most enigmatic are Tb$_2$Ti$_2$O$_7$ \cite{gardener1,gardener3,sato,Takatsu,yao,luo,lhotel,legl,peter,fennell} and Yb$_2$Ti$_2$O$_7$ \cite{hodges,ross09,ross11,chang}, which are under intense investigation.

In Tb$_2$Ti$_2$O$_7$, no long-range order has been found by muon-spin relaxation ($\mu$SR) down to 70\,mK and by neutron scattering to 50\,mK \cite{gardener1,gardener3}, two orders of magnitude lower than the absolute value of the crystal-field-subtracted Curie-Weiss temperature, about $-$13\,K or $-$7.0\,K \cite{gingras3,mirebeau}, raising the possibility that this magnet is a long-sought three-dimensional quantum spin liquid. Indeed, inelastic neutron scattering suggests a crossover from a thermally disordered paramagnet to a spin liquid at about 0.4\,K \cite{Takatsu}. But a sharp peak in specific heat, suggestive of long-range ordering, has been observed at 0.37\,K by a semi-adiabatic method \cite{sato}, although no peak has been detected by a relaxation method in a different sample \cite{yao}. On the other hand, a muon-spin-rotation ($\mu$SR) frequency shift in low magnetic fields, 20\,mT and 60\,mT, applied along the [110] direction suggests a transition at a lower temperature of about 150\,mK, a transition which has not been identified \cite{yao}. The static magnetic susceptibility of a zero-field-cooled polycrystalline sample in a 1\,mT field shows history dependence suggestive of spin-glass behavior below about 100\,mK, with an anomaly at about 70\,mK interpreted as the spin-glass transition from the high-temperature phase \cite{luo}.

Numerical diagonalization of a single-tetrahedron four-spin model predicts \cite{molavian,gingras} that
the zero-field ground state of Tb$_2$Ti$_2$O$_7$ is a quantum spin liquid, dubbed a quantum spin ice \cite{molavian,moessner}, which---with low increasing field along the [111] direction---gradually turns into a partially polarized state akin to the kagome-ice state \cite{matsuhira,tabata} of classical spin-ice magnets. At higher fields, $H$\,$\geq$\,82\,mT, it evolves into a ``three-in one-out" state, with three spins pointing into and one pointing out of every tetrahedron. The partially polarized state, the quantum kagome ice, will manifest itself as a magnetization plateau spanning from 18\,mT to 82\,mT, a feature detectable at and below 20\,mK.

Motivated by this prediction, we have made systematic measurements of the ac magnetic susceptibility of Tb$_2$Ti$_2$O$_7$
in zero magnetic field and fields up to 1.5\,T, applied along [111]. We find evidence for a new phase bounded in zero field at 140\,mK and, in the limit of zero temperature, at about 67\,mT. In this phase, Tb$_2$Ti$_2$O$_7$ exhibits frequency- and amplitude-dependent susceptibility and extremely slow spin dynamics, which is observed as the temperature is swept or the ac frequency is changed stepwise. At 16\,mK, the susceptibility indicates the presence of a weak magnetization plateau, adjacent to the low-field phase and extending to 0.59\,T---in qualitative, but not quantitative, agreement with the prediction.

The method of sample fabrication is similar to that of Ref.~\cite{gardener4}. A single crystal of Tb$_2$Ti$_2$O$_7$ was grown by the floating-zone technique at a rate of 2.5\,mm/hour in a 0.3\,MPa oxygen atmosphere.
The sample was approximately a square cuboid, 4\,mm\,$\times$\,2.4\,mm\,$\times$\,2.4\,mm, cut from the crystal with the long edges parallel to [111]. The setup for ac susceptibility measurements has been described elsewhere \cite{yin2,yin1,yu}. The ac field was applied along the [111] direction, as was the dc field. The ac frequencies ranged from 0.21\,Hz to 87.6\,Hz, and the ac-field amplitude, $H_{ac}$, was 0.20\,mT or 0.94\,mT. In addition, frequencies up to 389.6\,Hz were used to measure spin relaxation, and amplitudes ranging from 0.048\,mT to 1.2\,mT were used in a nonlinearity study described in the Supplemental Material \cite{suppl}. Immersing the sample in liquid $^3$He, which in turn was cooled by a dilution refrigerator, allowed us to make reliable measurements to at least 16\,mK, lower than in any previous experiment on Tb$_2$Ti$_2$O$_7$.

\begin{figure} [t]
\includegraphics[scale=0.26]{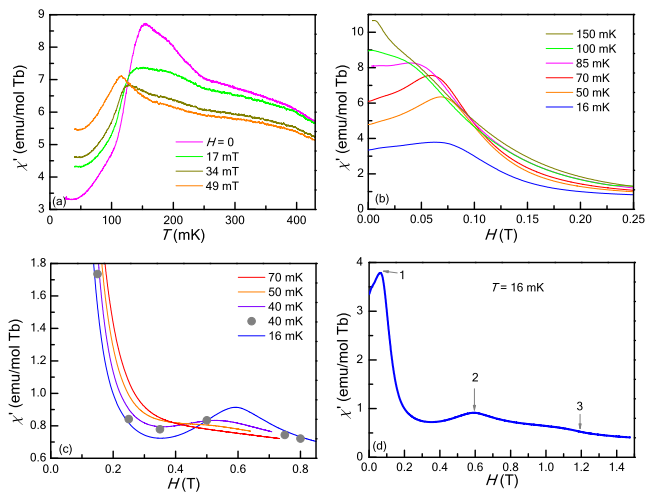}
\caption{(color online). (a) $\chi'$ as a function of temperature at various magnetic fields up to 49\,mT, measured during upward temperature sweeps at 0.6\,mK/min. The frequency and amplitude of the ac field were 0.21\,Hz and 0.94\,mT. (b) $\chi'$ as a function of magnetic field up to 0.25\,T at various temperatures. At and above 85\,mK, the frequency and amplitude of the ac field and the field-sweep rate were 87.6\,Hz, 0.94\,mT, and 4.2\,mT/min. For lower temperatures, they were reduced to 9.6\,Hz, 0.20\,mT, and 0.625\,mT/min. (With the ac-field amplitude of 0.94\,mT, the 100\,mK and 85\,mK data were taken in a slightly nonlinear regime, at least at $H$\,=\,0. See the Supplemental Material \cite{suppl}.) (c) The data at and below 70\,mK over a wider field range. Lines are field-sweep data. Circles are equilibrium data at 40\,mK. (d) The 16\,mK data up to 1.5\,T, arrows indicating anomalies discussed in the text. See the Supplemental Material \cite{suppl} for $\chi''$.}\label{fig1}
\end{figure}

Shown in Fig.~\ref{fig1} are the real part, $\chi'$, of the susceptibility as functions of temperature and magnetic field, for temperatures up to 430\,mK and fields up to 1.5\,T \cite{correction}. Because of the extremely slow relaxation of the spins, a subject discussed later in detail, the temperature was raised very slowly upward at a rate of 0.6\,mK/min in temperature sweeps. For field sweeps, the sweep rate at temperatures down to 86\,mK was 4.2\,mT/min, which was reduced to 0.625\,mT/min at lower temperatures in order to ensure a nearly equilibrium condition (see the Supplemental Material \cite{suppl}). As a result of the extremely slow relaxation, most of our data, even those at 0.21\,Hz, were taken in the adiabatic limit, not the isothermal limit.

At zero field, a peak appears in $\chi'$ at 154\,mK. When a small magnetic field, up to 49\,mT, is applied along [111], the peak becomes smaller and moves to lower temperatures, as shown in Fig.~\ref{fig1}(a).
In the field sweep at 16\,mK, a round peak occurs at 63\,mT, moving to lower fields as the temperature is raised, as shown in Fig.~\ref{fig1}(b). The peak vanishes completely between 85\,mK and 100\,mK. Most significantly, a second peak appears at 0.59\,T (the internal field $H_{int}$\,$\simeq$\,0.53\,T), as shown in Fig.~\ref{fig1}(c) \cite{demag}. This smaller, broader peak also moves to lower fields as the temperature is raised, becoming a very weak shoulder at 50\,mK and disappearing before the temperature reaches 70\,mK.
This peak is distinct from the weak anomaly observed by Legl \emph{et al}. \cite{legl} in the derivative of magnetization, $dM/dH$, at a higher field of 1.9\,T ($H_{int}$\,$\simeq$\,1.3\,T) at 70\,mK. It is very likely that that shoulder-like feature corresponds to the similar one in our data at about 1.2\,T ($H_{int}$\,$\simeq$\,1.0\,T), shown in Fig.~\ref{fig1}(d) \cite{demag}.

\begin{figure} [t]
\includegraphics[scale=0.3]{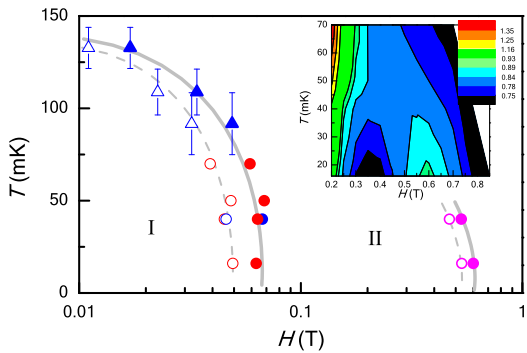}
\caption{(color online). \textit{H}-\textit{T} phase diagram determined from the $\chi'$ peak positions. Triangles are from the temperature sweeps, with peak positions in downward and upward sweeps averaged. Circles are from field sweeps. Red and magenta circles: from the data shown in Figs.~\ref{fig1}(b) and(c); blue circle: from 0.21\,Hz data, for which the ac amplitude and the field-sweep rate were 0.94\,mT and 0.51\,mT/min. Open symbols are the same data plotted against $H_{int}$. Lines are guides to the eye. Inset: contour plot of $\chi'$ in the $H$-$T$ plane in and near phase II, constructed from field and temperature sweeps at 9.6\,Hz.}\label{fig2} 
\end{figure}

The positions of the $\chi'$ peaks are summarized in Fig.~\ref{fig2}, which suggests the existence of two new phases---phase I, bounded at about 140\,mK in zero field and at about 70\,mT in the limit of zero temperature, and phase II, bounded at about 0.60\,T in the same limit. As shown in the inset, the upper boundary of phase II exists up to about 40\,mK but, unlike that of phase I, it gradually disappears at higher temperatures. This absence of a clear phase boundary to the high-temperature disordered phase is reminiscent of a classical kagome ice \cite{tabata}.

The field dependence of $\chi'$ at 16\,mK, with two peaks, indicates the presence of a weak magnetization plateau in phase II, between 63\,mT and 0.59\,T. This finding lends support to the quantum kagome-ice state predicted by the single-tetrahedron four-spin model and, by implication, to the underlying proposal \cite{molavian} that the low-field ground state of Tb$_2$Ti$_2$O$_7$ is a quantum spin ice. The temperature at which the second peak becomes evident, 40\,mK, is consistent with this prediction, strongly suggesting that the energy scale of the model is correct. The two critical fields, however, are higher than predicted, by a factor of about three and seven, respectively \cite{demag}, suggesting that the field dependence of the energy level of the three-in one-out state may be strongly affected by many-tetrahedron effects, which are absent from the model.

\begin{figure} [t]
\includegraphics[scale=0.32]{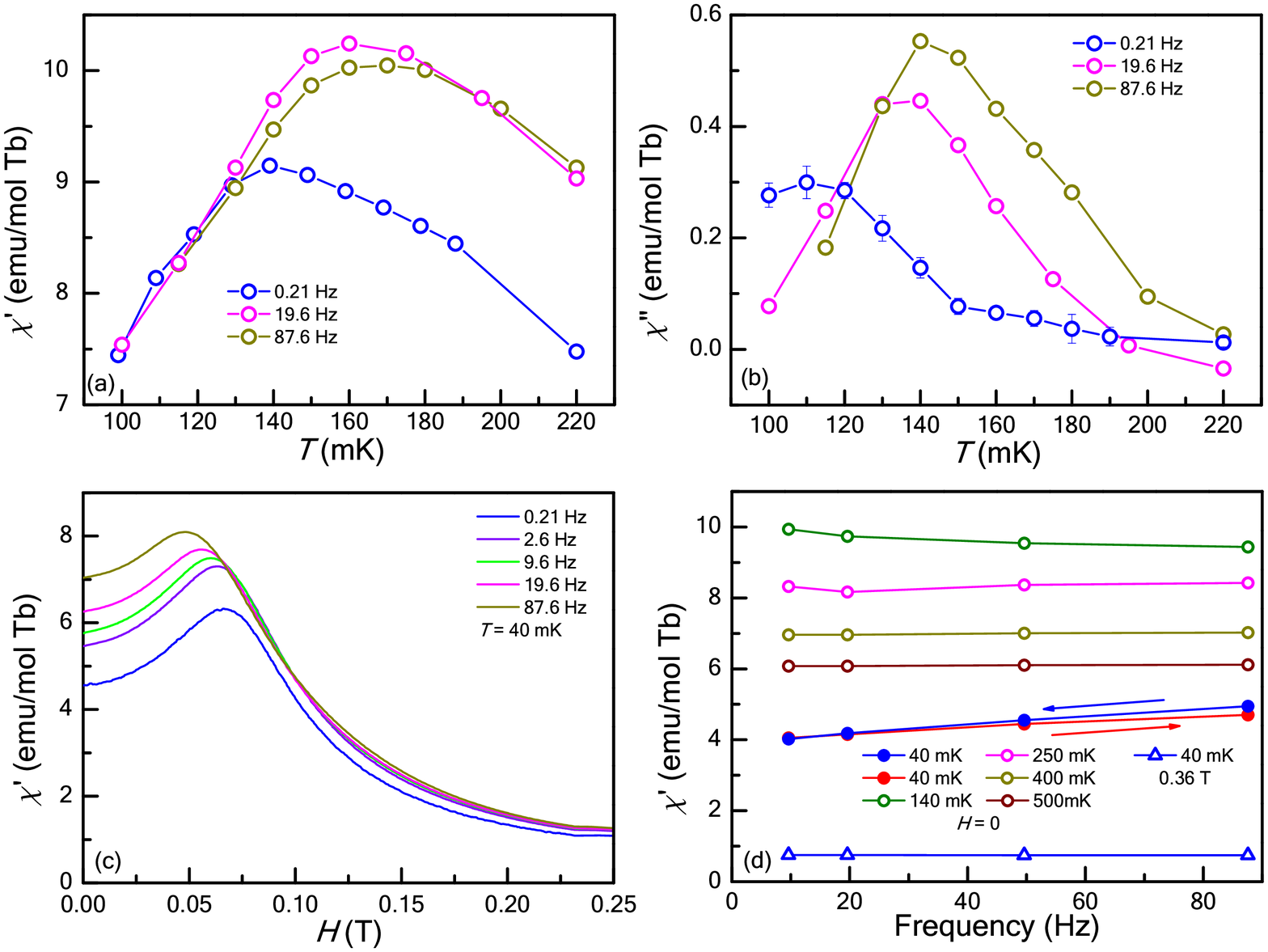}
\caption{(color online). Equilibrium $\chi'$ (a) and $\chi''$ (b) near the $\chi'$ peak in zero field. (c) $\chi'$ at 40\,mK as a function of magnetic field at various frequencies. The field was swept upward at 4.2\,mT/min. For all panels, the ac-field amplitude $H_{ac}$ was 0.94\,mT. Thus, in panels (a) and (b), the 87.6\,Hz data near 100\,mK were taken in a slightly nonlinear regime. Similarly, the 87.6\,Hz data in panel (c) were from a slightly nonlinear regime, at least at $H$\,=\,0 (see the Supplemental Material \cite{suppl}). (d) $\chi'$ at $H$\,=\,0 and 0.36\,T as a function of frequency at various temperatures. $H_{ac}$\ was 0.20\,mT. For the 40\,mK data at $H$\,=\,0, arrows indicate the directions of frequency steps, showing a weak hysteresis. Hysteresis is absent from all the other data. See the Supplemental Material \cite{suppl} for the imaginary part of the data shown in panel (c).}\label{fig3}
\end{figure}

Figures~\ref{fig3}(a) and \ref{fig3}(b) show the real and imaginary parts, $\chi'$ and $\chi''$, of the susceptibility near the $\chi'$ peak at zero field and at equilibrium, for three frequencies. At 0.21\,Hz, a round peak appears in $\chi'$ at about 140\,mK. With increasing frequency, the $\chi'$ peak becomes wider and moves to slightly higher temperatures. A peak also appears in $\chi''$ about 30\,mK below the $\chi'$ peak. Previous susceptibility measurements \cite{gardener3} have observed a similar frequency dependence, but with a broad peak at 16\,Hz appearing at about 250\,mK, clearly higher than 160\,mK at 19.6\,Hz for our sample. Measurements by other groups \cite{sato,lhotel} have also found a similar frequency dependence, with peaks in $\chi'$ and $\chi''$ occurring at temperatures closer to ours, for an ac field applied along [001] or [110].

The shift of a $\chi'$ peak to a higher temperature caused by increasing the frequency, observed in these and our measurements, is reminiscent of a spin-glass transition \cite{mydosh},
as is the appearance of a sharper peak in $\chi''$. Moreover, we find that the $\chi'$-peak shift per decade frequency is $\Delta T_{p}/(T_{p}\log f)$\,=\,0.075, where $T_{p}$ is the peak temperature and $f$ the frequency, consistent with 0.06--0.08 from more extensive measurements \cite{lhotel} and similar to that of a spin-glass transition in an insulator \cite{mydosh}. However, the downward shift of a $\chi'$ peak with increasing field (Fig.~\ref{fig1}(a)) is not found in a canonical spin glass, where application of a field instead moves the peak slightly upward in temperature \cite{mydosh}.

The peak position in field sweeps also depends on the frequency, moving to lower fields with increasing frequency, as shown in Fig.~\ref{fig3}(c). Therefore, the boundary of phase I drawn in Fig.~\ref{fig2} is frequency-dependent. Moreover, $\chi'$ at zero field increases linearly with increasing frequency and exhibits slight hysteresis, as shown in Fig.~\ref{fig3}(d). This frequency dependence is much weaker at and above 140\,mK, outside phase I, and is absent at 0.36\,T, well into phase II.

\begin{figure} [t]
\includegraphics[scale=0.3]{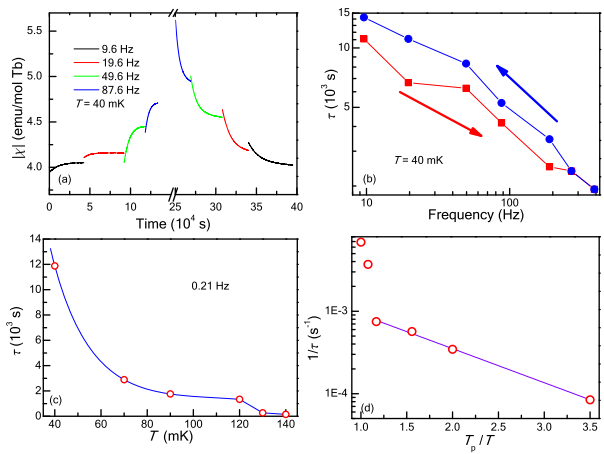}
\caption{(color online). (a) $|\chi|$ as a function of time at 40\,mK in zero field. Before $t$\,=\,0, the ac frequency was held at 2.6\,Hz for 12\,hours until equilibrium was reached. Thereafter, the frequency was changed stepwise, each time after new equilibrium was reached. For clarity, relaxation curves at 189.6\,Hz, 269.6\,Hz, and 389.6\,Hz---between the two 87.6\,Hz curves---are not shown.
(b) Relaxation time $\tau$ as a function of frequency, extracted from the data. Arrows indicate the frequency-step sequence.
(c) $\tau$ \emph{vs} temperature. At each temperature, the frequency was first held at 389.6\,Hz until equilibrium was reached, then stepped down to 0.21\,Hz, the frequency at which the measurements were made. (d) 1/$\tau$ \emph{vs} $T_{p}$/$T$, where $T_p$\,=\,140\,mK. For all the panels, $H_{ac}$\,=\,0.20\,mT.}\label{fig4}
\end{figure}

Although its mechanism is not understood, this peculiar frequency dependence of $\chi'$ allows us to accurately measure the relaxation time of spins by stepping the frequency, which---unlike temperature---can be changed nearly instantaneously. Figure~\ref{fig4}(a) shows the total magnitude, $|\chi|$, of the susceptibility at 40\,mK at zero field, as the frequency was changed stepwise between 9.6\,Hz and 389.6\,Hz, first upward then downward. $|\chi|$, instead of $\chi'$ alone, is fitted to a stretched exponential $\exp[-(t/\tau)^\alpha]$ to extract the relaxation time $\tau$, because both $\chi'$ and $\chi''$ evolve with time after a frequency step. The stretching exponent $\alpha$ ranges from 0.94 to 1.10, indicating that the relaxation is nearly exponential. The relaxation time is a strong function of frequency at this temperature, as shown in Fig.~\ref{fig4}(b).
It also depends strongly on temperature, as Fig.~\ref{fig4}(c) shows, with an abrupt rise within a narrow temperature interval between 120\,mK and 130\,mK, where $\chi''$ rises sharply (see Fig.~\ref{fig3}(b)). The relaxation time measured at 0.21\,Hz up to 120\,mK can be expressed by an empirical formula for spin glasses \cite{hoogerbeets}, 1/$\tau$ $\sim$ $\exp[-a(T_p/T)]$ with $a$\,=\,0.41 and $T_p$\,=\,140\,mK, as shown in Fig. \ref{fig4}(d). Note that even at 140\,mK, the relaxation rate, 1/$\tau$, is as small as 7.1\,$\times$\,10$^{-3}$\,s$^{-1}$, two orders of magnitude smaller than even our lowest measuring frequency, indicating that much of the susceptibility we have measured is in the adiabatic limit, not the isothermal limit.

A similar temperature dependence has been found in a spin glass \cite{svedlindh,lundgren}. Unlike in our case, however, the relaxation is logarithmic in time, as predicted by the droplet scaling theory \cite{fisher}, not nearly exponential. To our knowledge, relaxation rates much smaller than measuring frequencies have not been observed in a spin glass.

In summary, we have constructed, from ac magnetic susceptibility, a low-temperature magnetic phase diagram of Tb$_2$Ti$_2$O$_7$ in low magnetic fields applied along [111], uncovering two phases. The field dependence of the susceptibility at 16\,mK and 40\,mK indicates the presence of a weak magnetization plateau in support of theory \cite{molavian,gingras}, suggesting that phase II is a quantum kagome ice. In view of the underlying proposal of the theory, phase I is likely to be a quantum spin ice. The frequency dependence of the susceptibility-peak position at zero field is similar to that of a spin glass, as is the temperature dependence of the relaxation time. But the field dependence of the peak position and the time dependence of the relaxation, as well as the frequency and amplitude dependence of $\chi'$ at 40~mK, suggest
that phase I is not a spin glass.

In addition to our work, three very recent studies have investigated Tb$_2$Ti$_2$O$_7$ in magnetic fields applied along [111], also motivated by the prediction on the low-field ground state by the theory. Lhotel \emph{et al.} \cite{lhotel} measured the magnetization up to 8\,T and ac susceptibilities at zero field, and found no evidence for a magnetization plateau. However, the lowest temperature of the experiment was 80\,mK, insufficient to detect the weak plateau indicated by our susceptibility result. Legl \emph{et al.}\,\cite{legl} also measured the magnetization, to a lower temperature of 43\,mK, yet finding no evidence for a plateau. In particular, the derivative $dM/dH$ exhibits no anomaly that corresponds to the second $\chi'$ peak we have observed at and below 40\,mK. Our results shown in Fig.~\ref{fig1}(c) suggest, however, that 43\,mK may not be cold enough to clearly detect such an anomaly. Moreover, the field-sweep rate employed by Legl \emph{et al.} was four times faster than the rate we have found at 40\,mK to completely wipe out the peak (see the Supplemental Material \cite{suppl}). The low-field low-temperature phase diagram proposed by these authors also differs substantially from ours. However, at least part of the discrepancy may come from the much faster temperature-sweep rate---5\,mK/min as opposed to our 0.6\,mK/min---used by Legl \emph{et al.}. Baker \emph{et al.} \cite{peter} measured $\mu$SR to 25\,mK and ac susceptibility to 68\,mK. They have observed anomalies in the $\mu$SR relaxation time at 15\,mT and $\sim$60\,mT, at and below 50\,mK. The susceptibility $\chi'$ at 68\,mK and 100\,mK, measured at 50\,Hz and 500\,Hz, is similar to ours but shows a broad peak at about 15\,mT, lower than the positions of the peak we have found at 9.6\,Hz and 87.6\,Hz at similar temperatures (Fig.~\ref{fig1}(b)). The temperature and field ranges of the experiment were not sufficient to observe the second peak, the key signature of the weak magnetization plateau.

\begin{acknowledgments}
We thank C.\,P.~Aoyama for making SQUID-magnetometer measurements for calibration and P.\,J.\ Baker and H.\,D.\ Zhou for helpful discussions. The experiment was performed at the NHMFL High B/T Facility, which is supported by NSF Grant DMR 0654118 and by the State of Florida. Q.\,J.\,L. and X.\,F.\,S. acknowledge supports from the National Natural Science Foundation of China, the National Basic Research Program of China (Grant Nos. 2009CB929502 and 2011CBA00111), and the Fundamental Research Funds for the Central Universities (Program No. WK2340000035).

\end{acknowledgments}


\begin{thebibliography}{}
\bibitem{demag}
Using an average demagnetization factor for our sample, 0.23, we estimate the internal fields, $H_{int}$, of the three anomalies marked in Fig.~\ref{fig1}(d) to be 48\,mT, 0.53\,T, and 1.0\,T. This should be compared with the applied fields, 63\,mT, 0.59\,T, and 1.2\,T.
\end{thebibliography}
\end{document}


\preprint{}
\title{Low-Temperature Low-Field Phases of the Pyrochlore Quantum Magnet Tb$_2$Ti$_2$O$_7$ (Supplemental Material)}
\author{L. Yin}
\email{yin@phys.ufl.edu}
\author{J. S. Xia}
\author{Y. Takano}
\author{N. S. Sullivan}
\affiliation{National High Magnetic Field Laboratory and Department of Physics, University of Florida, Gainesville, FL 32611, USA}
\author{Q. J. Li}
\author{X. F. Sun}
\email{xfsun@ustc.edu.cn}
\affiliation{Hefei National Laboratory for Physical Sciences at Microscale, University of Science and Technology of China, Hefei, Anhui 230026, People's Republic of China}

\maketitle

{\it Effect of slow relaxation.}
Spin relaxation in Tb$_2$Ti$_2$O$_7$ slows down dramatically in phase I as the temperature decreases, as shown in Fig.~4(c) of the accompanying Letter. Therefore, great care must be exercised to ensure equilibrium when measurements are made while sweeping the temperature or the magnetic field.

\begin{figure} [H]
\begin{center}
\includegraphics[scale=0.36]{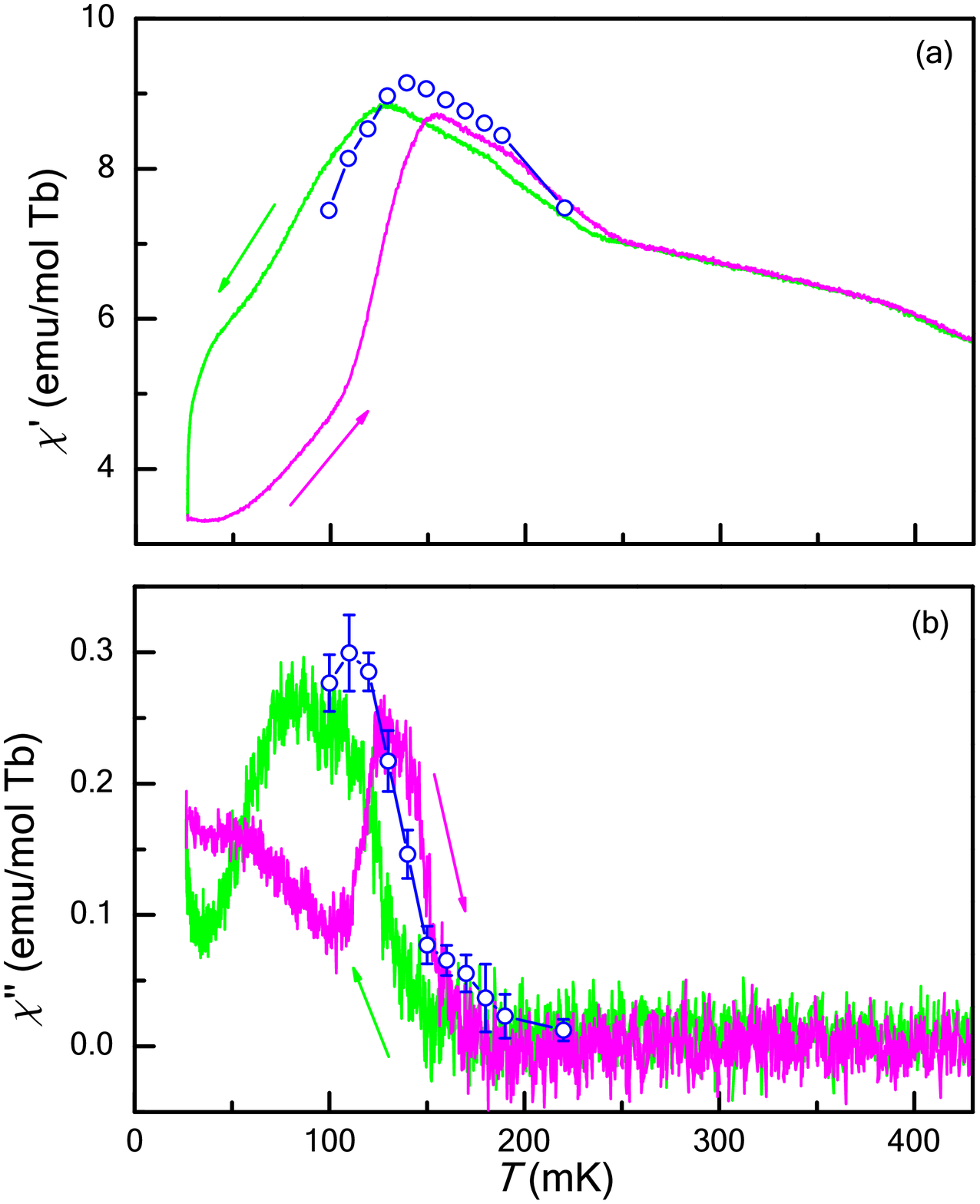}
\end{center}
\caption{(color online). (a) Real part, $\chi'$, and (b) imaginary part, $\chi''$, of the ac susceptibility at 0.21\,Hz as a function of temperature at zero field, with arrows indicating the directions of temperature sweeps. Blue circles are the equilibrium data shown in Figs.~3(a) and 3(b). The ac-field amplitude was $H_{ac}$\,=\,0.94\,mT.}\label{fig1sup}
\end{figure}

The effect of the slow relaxation on ac susceptibility measured during temperature sweeps was studied at zero field with an ac frequency of 0.21\,Hz and an amplitude of 0.94\,mT. Starting from 500\,mK, the temperature was lowered to 25\,mK, where it was held for eight hours, then raised back to 500\,mK. The temperature regulator was set at a sweep rate of 0.6~mK/min, but the actual cooling rate gradually decreased below 100\,mK, limited by the cooling power of the dilution refrigerator; for instance, at 70\,mK it was 0.2\,mK/min.

Below about 150\,mK,
the cooling and warming curves of both $\chi'$ and $\chi''$ of the ac susceptibility deviate strongly from each other, as shown in Fig.~\ref{fig1sup} \cite{correction}. But the equilibrium data taken near the $\chi'$ peak shows no hysteresis within our resolution, indicating that the hysteresis arises from slow spin relaxation. Indeed, at the lowest temperature, 25\,mK, the relaxation time of $\chi'$ was as long as 6.8\,hours. Because the warming rate was constant, unlike the cooling rate, only data taken upon warming are shown in Fig.~1(a) of the accompanying Letter. The data shown in Fig.~\ref{fig1sup} suggest, however, that the average of the $\chi'$-peak positions in the cooling and warming curves is very close to that of the equilibrium data. In Fig.~2 of the Letter, such averages are used in part to construct the phase diagram.

\begin{figure}[H]
\begin{center}
\includegraphics[scale=0.32]{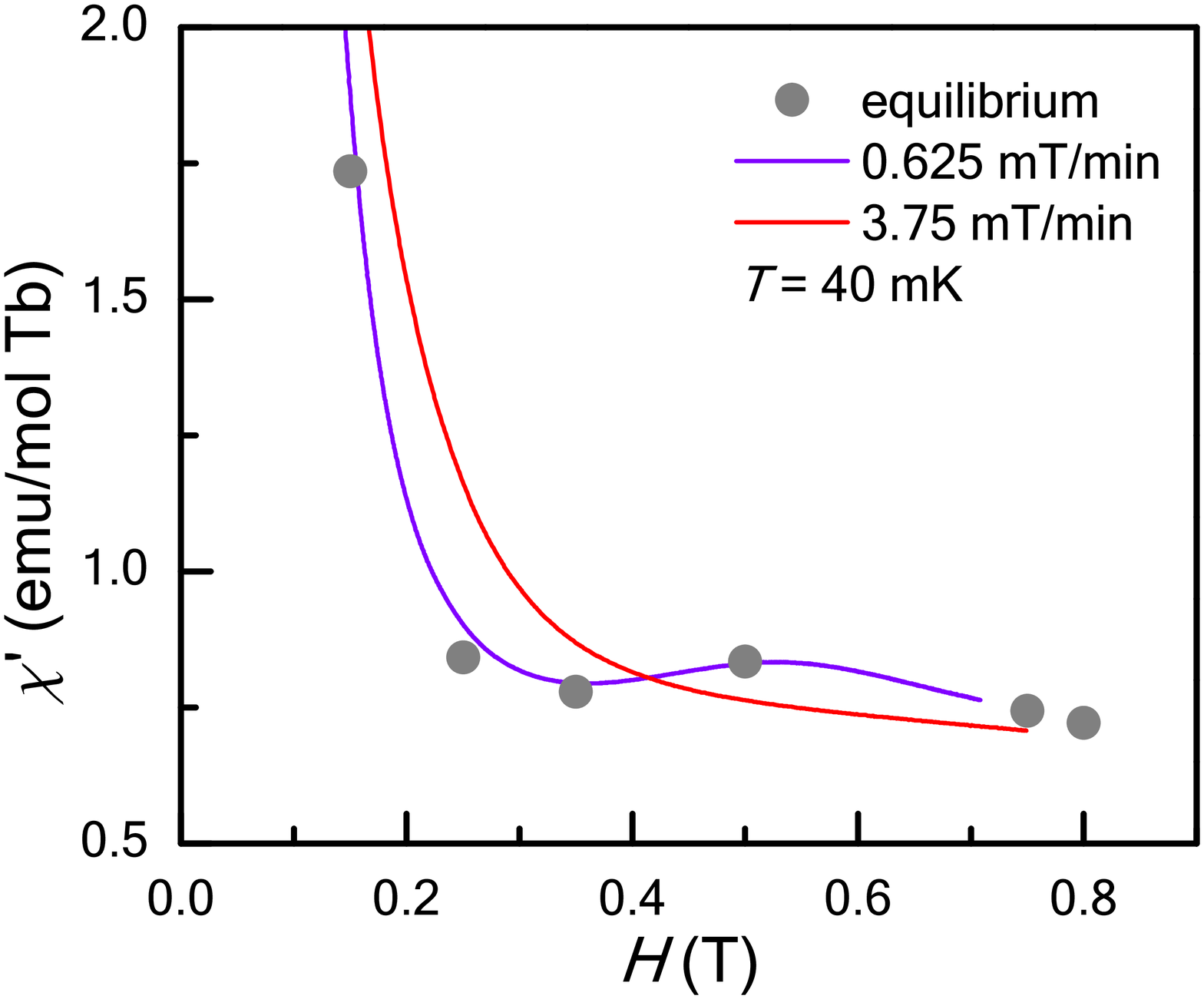}
\end{center}
\caption{(color online). $\chi'$ as a function of magnetic field at 40\,mK, for two field-sweep rates in comparison with equilibrium data. The magnetic field, applied along the [111] direction, was swept upward. The ac frequency was 9.6\,Hz, and $H_{ac}$ was 0.20\,mT. The equilibrium data and the 0.625\,mT/min data are shown also in Fig.~1(c).}\label{S2}
\end{figure}

Great care must be exercised also when data are taken while sweeping the field. The effect of the slow relaxation on field-sweep data was studied at 40\,mK with an ac frequency of 9.6\,Hz and an amplitude of 0.20\,mT. We have found that even a sweep rate as low as 3.75\,mT/min, a quarter of the rate used by Legl.~\emph{et al.} \cite{legl}, completely wipes out the 0.52\,T peak in $\chi'$, whereas a field sweep at 0.625\,mT/min reveals the peak in agreement with equilibrium data taken at fixed fields, as shown in Fig.~\ref{S2}. For this reason, all the field-sweep data shown in Fig.~1(c) of the accompanying Letter were taken with a sweep rate of 0.625\,mT/min.

\begin{figure} [H]
\begin{center}
\includegraphics[scale=0.31]{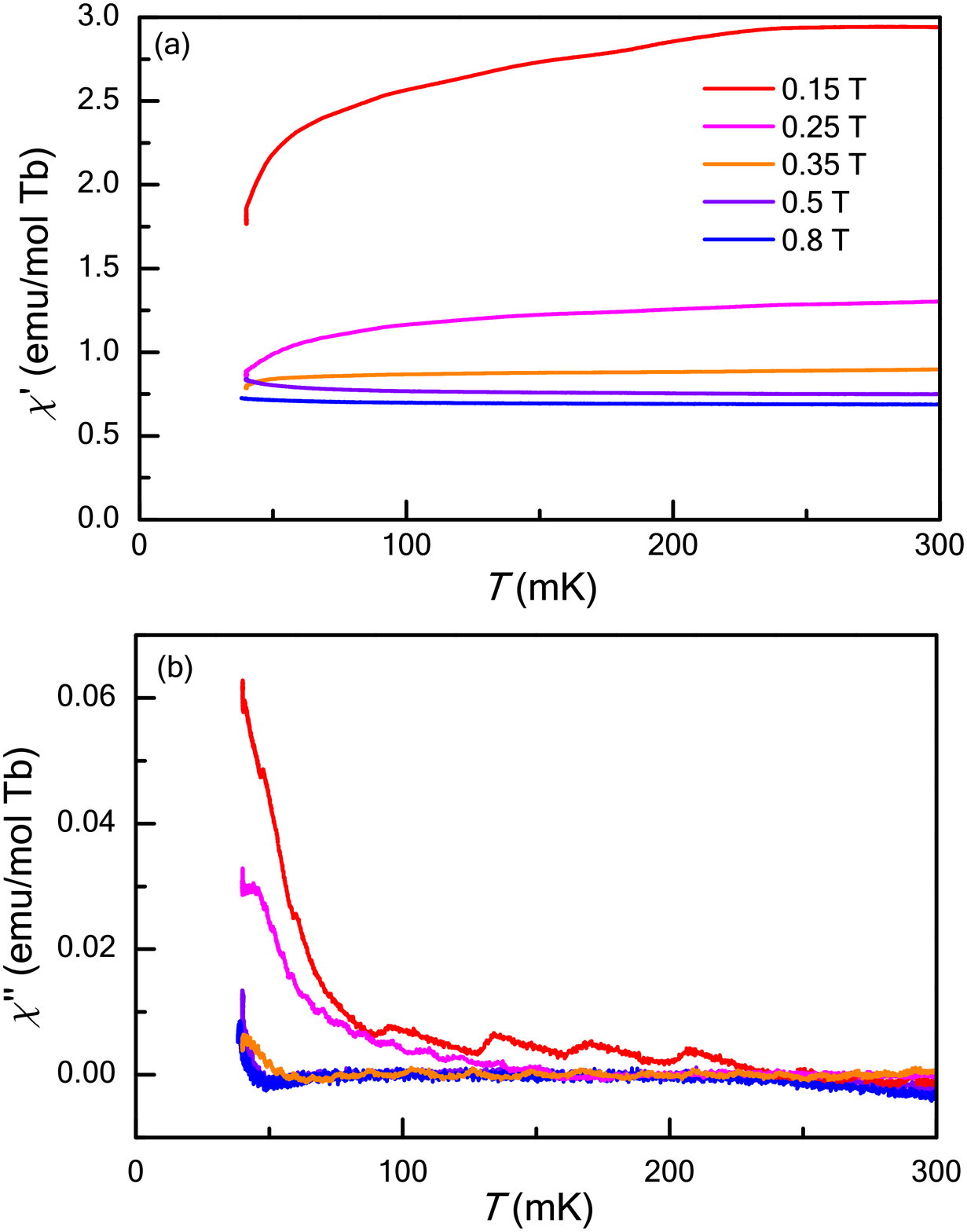}
\end{center}
\caption{(color online). $\chi'$ and $\chi''$ as a function of temperature at various fields. The frequency and amplitude of the ac field were 9.6\,Hz and 0.20\,mT. The temperature was raised at a rate of 0.7\,mK/min.}\label{S5}
\end{figure}

{\it Temperature sweeps at and above 0.15\,T.}
Figure~\ref{S5} shows $\chi'$ and $\chi''$ in the field region of phase II and, for comparison, at 0.8\,T, which is slightly above phase II. No anomaly appears in these temperature-sweep data, indicating that the boundary of phase II is open, unlike the closed boundary of phase I. $\chi''$ is very small at these fields; as a result, it sometimes picks up semi-periodic noise, which may come from the air conditioning of the laboratory.

\begin{figure} [H]
\begin{center}
\includegraphics[scale=0.3]{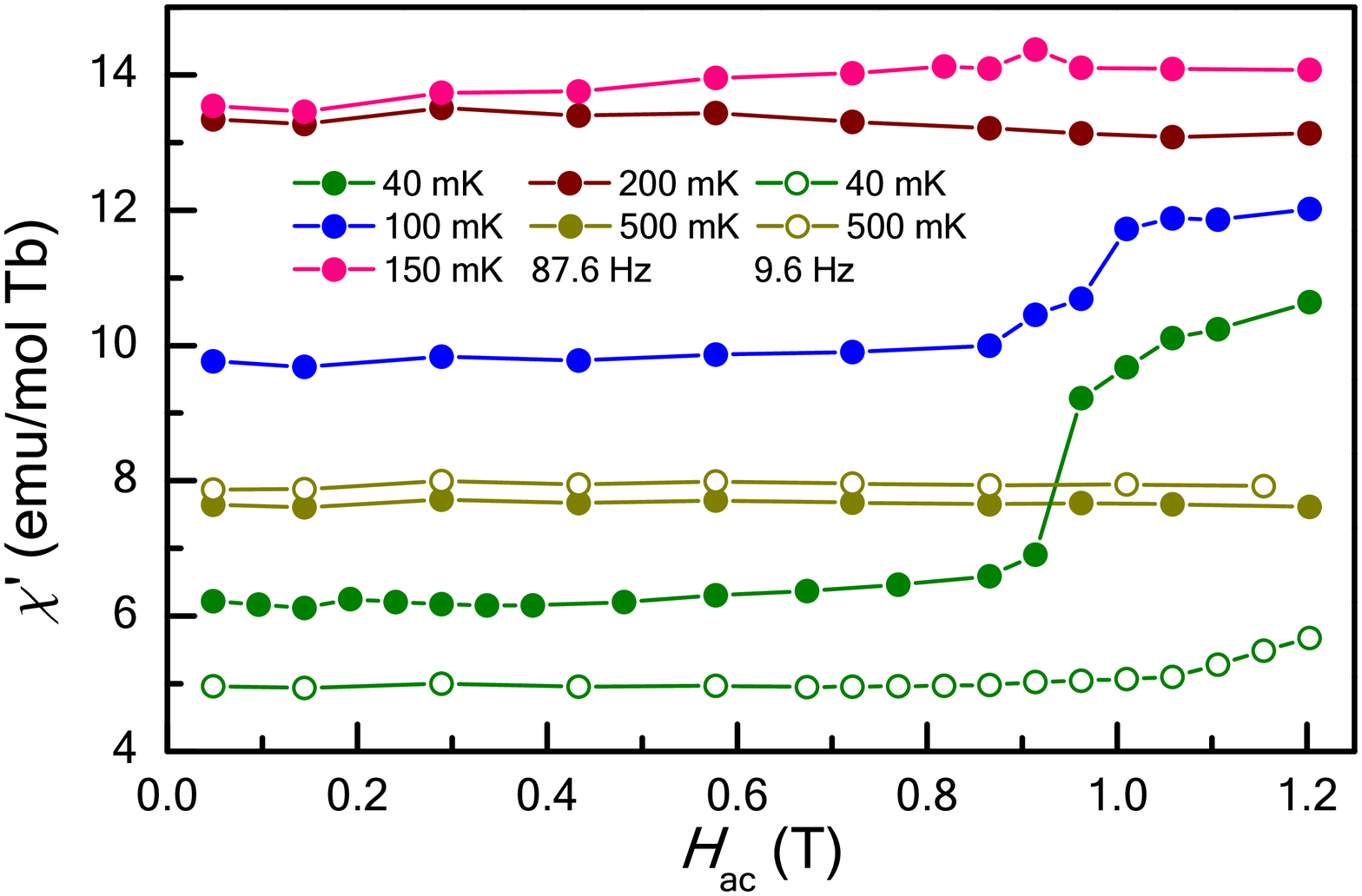}
\end{center}
\caption{(color online). Dependence of $\chi'$ at $H$\,=\,0 on the ac-field amplitude, $H_{ac}$, at 87.6\,Hz (solid circles) and 9.6\,Hz (open circles).}\label{S3}
\end{figure}

{\it Amplitude dependence of susceptibility in phase I.}
In zero field, and at and blow 100\,mK, $\chi'$ depends on the ac-field amplitude $H_{ac}$. The amplitude dependence has an onset, as shown in Fig.~\ref{S3}. At 87.6\,Hz, the onset occurs at $H_{ac}$\,$\simeq$\,0.9\,mT, above which $\chi'$ increases with increasing amplitude, whereas the dependence is weaker at 9.6~Hz, with the onset moving to $H_{ac}$\,$\simeq$\,1.1\,mT. At and above 150\,mK, outside phase I, we do not see an amplitude dependence even at 87.6\,Hz, at least up to $H_{ac}$\,=\,1.2\,mT. The mechanism of this peculiar $H_{ac}$ dependence is not understood.

All the data shown in the accompanying Letter were taken in the amplitude-independent, linear regime---except for the 100\,mK and 85\,mK data in Fig.~1(b), the 87.6\,Hz data near 100\,mK in Figs.~3(a) and 3(b), and the 87.6\,Hz data in Fig.~3(c). Those data were taken with $H_{ac}$\,=\,0.94\,mT, which is in a slightly nonlinear regime at least at $H$\,=\,0.

\begin{figure} [H]
\begin{center}
\includegraphics[scale=0.34]{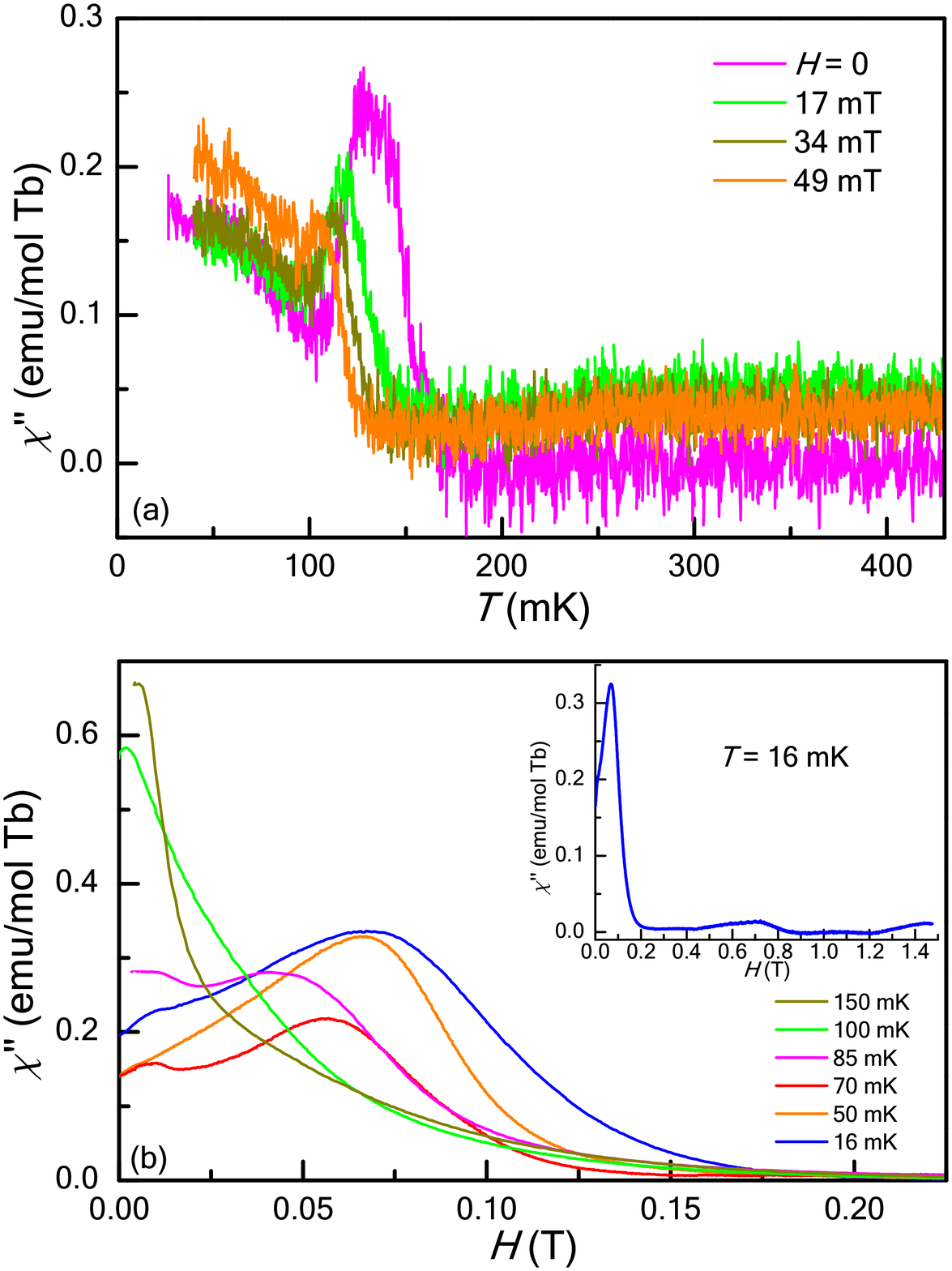}
\end{center}
\caption{(color online). (a) Imaginary part, $\chi''$, of the ac susceptibility whose real part is shown in Fig.~1(a) of the accompanying Letter. (b) $\chi''$ of the susceptibility whose real part is shown in Figs.~1(b) and 1(d). The small hump at a field below 0.02\,T comes from the background without the sample.}\label{S4}
\end{figure}

\begin{figure} [H]
\begin{center}
\includegraphics[scale=0.28]{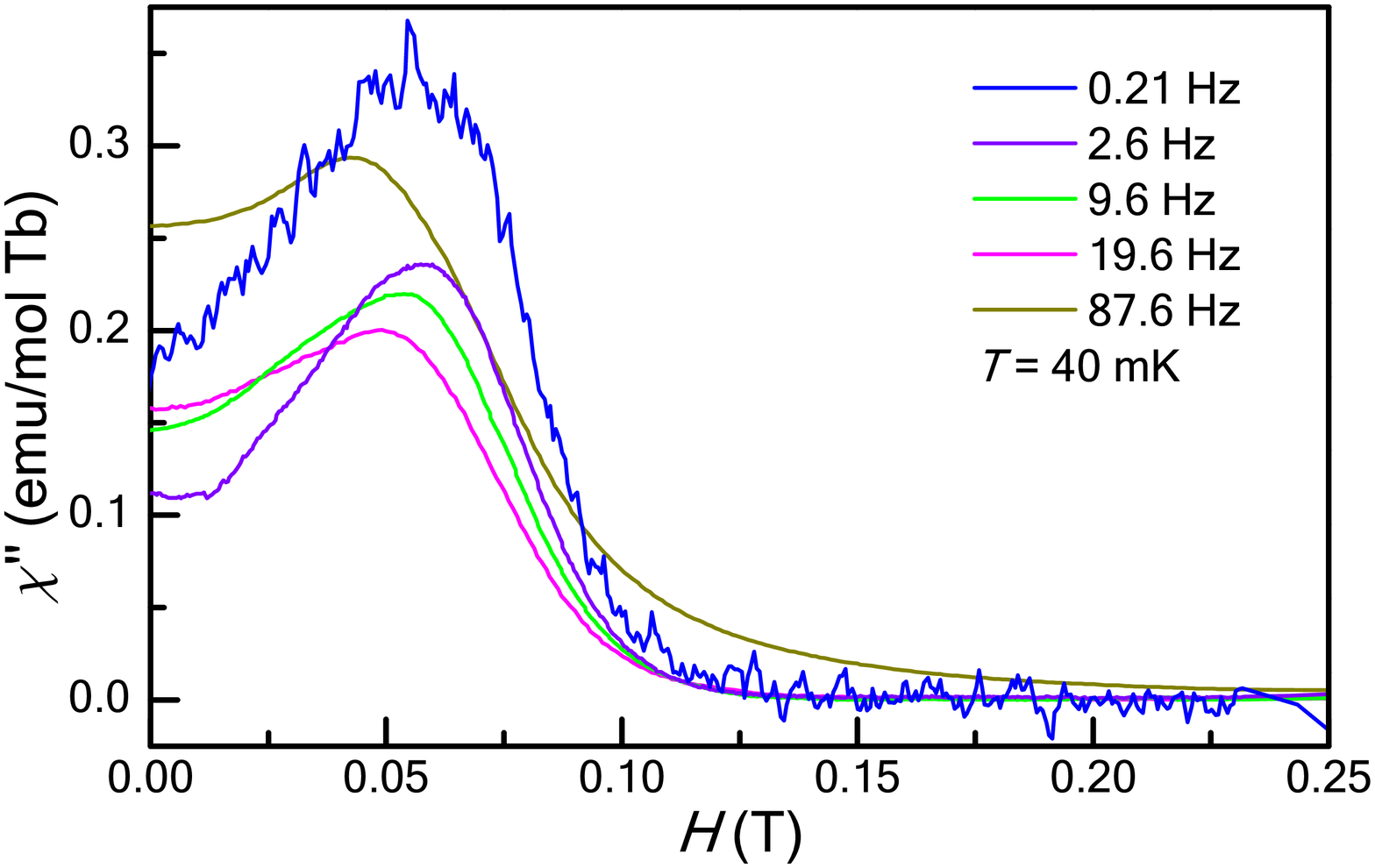}
\end{center}
\caption{(color online). Imaginary part, $\chi''$, of the ac susceptibility whose real part is shown in Fig.~3(c).}\label{S6}
\end{figure}

{\it Imaginary part of susceptibility.}
For completeness, we present the imaginary part, $\chi''$, of the susceptibility data whose real part is shown in Figs.~1(a), 1(b), 1(d), and 3(c) of the accompanying Letter. As shown in Fig.~\ref{S4}(a), the peak in $\chi''$ is sharper than the $\chi'$ peak (see Fig.~1(a)) and appears about 28\,mK below the latter. With increasing field, the $\chi''$ peak moves to lower temperatures along with the $\chi'$ peak, while at the same time becoming smaller. In field sweeps, the $\chi''$ peak is roughly as broad as the $\chi'$ peak, appearing about 6\,mT above the latter. It moves to lower fields along with the latter, as the temperature is raised, as shown in Fig.~\ref{S4}(b). $\chi''$ is negligible at fields above about 0.2\,T. As shown in Fig.~\ref{S6}, the effect of raising the ac frequency is qualitatively similar to that of raising the temperature; with increasing frequency, the broad $\chi''$ peak moves to lower fields along with the $\chi'$ peak.

